\documentclass[floatfix,aps,superscriptaddress,prd,amsmath,nofootinbib,preprintnumbers,onecolumn]{revtex4}
\usepackage{diagbox}
\usepackage{graphicx}
\usepackage{bm}
\usepackage{float}
\usepackage[
colorlinks=true,        
citecolor=blue,         
linkcolor=blue,         
urlcolor=blue           
]{hyperref}             

\newcommand{\nc}{\newcommand*}
\nc{\Om}{\Omega}
\nc{\ogw}{\Omega_{\mathrm{GW}}}
\nc{\rd}{\mathrm{d}}
\nc{\eg}{\textit{e.g.~}}
\nc{\red}[1]{\textcolor{red}{#1}}
\nc{\lvc}{LIGO/Virgo} 

\def\({\left(}
\def\){\right)}
\def\[{\left[}
\def\]{\right]}

\def\e{\begin{equation}}
\def\q{\end{equation}}
\def\m{\begin{eqnarray}}
\def\n{\end{eqnarray}}

\begin{document}

\title{Measuring the speed of scalar induced gravitational waves from observations}

\author{Jun Li}
\email{lijun@qust.edu.cn}
\affiliation{School of Mathematics and Physics,
    Qingdao University of Science and Technology,\\
    Qingdao 266061, China}
\affiliation{CAS Key Laboratory of Theoretical Physics,\\
    Institute of Theoretical Physics, Chinese Academy of Sciences,\\
    Beijing 100190, China}

\author{Guang-Hai Guo}
\email{ghguo@qust.edu.cn}
\affiliation{School of Mathematics and Physics,
    Qingdao University of Science and Technology,\\
    Qingdao 266061, China}

\date{\today}

\begin{abstract}
We investigate the scalar induced gravitational waves which propagate with a speed different from the speed of light. First, we analytically calculate the expression of the power spectrum of the scalar induced gravitational waves which is based on the speed and the spectrum of the primordial curvature perturbations. Then, we discuss several scalar power spectra and obtain corresponding fractional energy density, such as the monochromatic power spectrum, the scale invariant power spectrum and the power-law power spectrum. Finally, we constrain the scalar induced gravitational waves and evaluate the signatures of the speed from the combination of CMB+BAO and gravitational waves observations. The numerical results are obvious to reveal the influence of speed of scalar induced gravitational waves.

\end{abstract}

\maketitle

\section{introduction}
Since the discovery of temperature anisotropies in the cosmic microwave background (CMB), it becomes significant to study the early universe and the cosmological evolution \cite{Riotto:2002yw,Cabella:2004mk,Katsuki:1995ai,Zaldarriaga:1996xe}. We are interested in scalar perturbations to the metric since these couple to the density of matter and radiation and ultimately are responsible for most of the inhomogeneities and anisotropies in the universe. Inflation also generates tensor fluctuations in the metric, so-called gravitational waves. The scalar and tensor modes are decoupled in the first order of perturbation. Primordial gravitational waves are not couple to the density and so are not responsible for the large-scale structure of the universe, but they do induce fluctuations in the CMB \cite{Ezquiaga:2021ler,Saikawa:2018rcs,Campeti:2020xwn,Cai:2016ldn,Giare:2020vss,Brax:2017pzt,Dubovsky:2009xk,Lin:2016gve,Cai:2020ovp,Li:2017cds,Li:2018iwg,Li:2019efi,Li:2019vlb,Li:2021scb,Li:2021nqa}. The theoretical first-order approximation from perturbations has been revealed through the CMB observations which include the Wilkinson Microwave Anisotropy Probe (WMAP) \cite{WMAP:2008lyn}, the Planck satellite \cite{Planck:2018vyg}, the BICEP and Keck array (BK) \cite{BICEP:2021xfz}, and the Baryon Acoustic Oscillation (BAO) \cite{Beutler:2011hx,Ross:2014qpa,BOSS:2016wmc}. 

However, the rapid developments of the cosmological observations inspire us considering the deviation from the first-order approximation. The curvature perturbations couple to the tensor perturbations at second order which produce the scalar induced gravitational waves in the radiation dominated era \cite{Fu:2019vqc,Pi:2020otn,Hajkarim:2019nbx,Domenech:2020kqm,Yi:2020kmq,Tomikawa:2019tvi,Inomata:2019yww,Hwang:2017oxa,Domenech:2020xin,Yuan:2020iwf,Yuan:2019udt,Jinno:2013xqa,Chen:2019xse,Bartolo:2018rku,Tada:2019amh,Espinosa:2018eve,Di:2017ndc,Jin:2023wri,You:2023rmn,Orlofsky:2016vbd,Alabidi:2013lya,Osano:2006ew,Noh:2004bc,Matarrese:1997ay,Giovannini:2010tk,Xu:2019bdp,Unal:2018yaa,Cai:2018dig,Cai:2019jah,Cai:2019elf,Cai:2019amo,Assadullahi:2009nf,Assadullahi:2009jc,Inomata:2019zqy,Inomata:2019ivs,Inomata:2018epa,Yuan:2019fwv,Yuan:2019wwo,Zhou:2020kkf,Alabidi:2012ex,Kohri:2018awv,Lu:2019sti,Li:2022avp,Ananda:2006af,Baumann:2007zm,Inomata:2016rbd,Li:2021uvn}. The gravitational waves detections provide the latest way to find scalar induced gravitational waves which include Laser Interferometer Gravitational-wave Observatory (LIGO) and Virgo detector \cite{Thrane:2013oya,LIGOScientific:2016jlg,LIGOScientific:2019vic}, Laser Interferometer Space Antenna (LISA) detector \cite{Caprini:2015zlo}, International Pulsar Timing Array (IPTA) \cite{Verbiest:2016vem}, Five-hundred-meter Aperture Spherical radio Telescope (FAST) \cite{Nan:2011um, Kuroda:2015owv} and Square Kilometer Array (SKA) \cite{Kuroda:2015owv}. IPTA is the combination of three Pulsar Timing Array (PTA) projects \cite{Hellings:1983fr}, namely European Pulsar Timing Array (EPTA) \cite{EPTA:2016ndq}, Parkes Pulsar Timing Array (PPTA) \cite{Hobbs:2013aka} and North American Observatory for Gravitational Waves (NANOGrav) \cite{McLaughlin:2013ira}. If there is no detection of scalar induced gravitational waves from SKA project, the scalar amplitude $A_s$ and  the spectral index $n_s$ of the power-law spectrum are smaller than the constraints from CMB+BAO data which have introduced in our previous work \cite{Li:2022avp}. The upper limits of scalar induced gravitational waves from SKA project have given better constraints, while the influence of nontrivial speed of scalar induced gravitational waves is also significant.  

The nontrivial speed of gravitational waves is a fundamental issue on the propagation of gravitational waves. In general relativity, gravitational waves propagate at the speed of light. When gravitational waves propagate with a speed different from the speed of light at large scales, this scenario would arise a variety of modified gravity theories. Probing the speed of gravitational waves is an important way to explore modified gravity theories and underlying new physics. The nontrivial speed has been considered through primordial gravitational waves \cite{Brax:2017pzt,Lin:2016gve,Giare:2020vss,Cai:2020ovp,Cai:2016ldn,Ezquiaga:2021ler}, while the research on the speed of scalar induced gravitational waves are less which need more attention. Although the scalar induced gravitational waves are suppressed by the square of curvature perturbations, but they can compare with primordial gravitational waves if the curvature perturbations are large enough. It is worth looking forward to the signatures of the speed of scalar induced gravitational waves.

In this paper, we investigate the scalar induced gravitational waves which propagate with a speed different from the speed of light. In section 2, we analytically calculate the expression of the power spectrum of the scalar induced gravitational waves with nontrivial speed, and discuss several scalar power spectra, such as the monochromatic power spectrum, the scale invariant power spectrum and the power-law power spectrum. In section 3, we constrain the scalar induced gravitational waves and evaluate the signatures of the speed from the combination of CMB+BAO and gravitational waves observations. A brief summary will be given in section 4.

\section{the scalar induced gravitational waves}
In the conformal Newtonian gauge, the metric about the Friedmann-Robert-Walker background is taken as
\e
\mathrm{d}s^2=a^2\left\{-(1+2\Phi)\mathrm{d}\eta^2+\left[(1-2\Phi)\delta_{ij}+\frac{h_{ij}}{2}\right]\mathrm{d}x^i\mathrm{d}x^j \right\},      \label{metric}
\q
where $\eta$ is the conformal time, $a(\eta)$ is the scale factor, $\Phi$ is the scalar perturbation and $h_{ij}$ is the tensor perturbation. We neglect the vector perturbation, the first-order gravitational waves and the
anisotropic stress.
In the Fourier space, the tensor perturbation $h_{ij}$ is
\e
h_{ij}(\eta,\mathbf{x})=\int\frac{\mathrm{d}^3k}{(2\pi)^{3/2}}\Bigg(e_{ij}^{+}(\mathbf{k})h_{\mathbf{k}}^+(\eta)+e_{ij}^{\times}(\mathbf{k})h_{\mathbf{k}}^{\times}(\eta)\Bigg)e^{i\mathbf{k}\cdot\mathbf{x}},
\q
where the plus and cross polarization tensors are
\e
e_{ij}^{+}(\mathbf{k})=\frac{1}{\sqrt{2}}\Bigg(e_i(\mathbf{k})e_j(\mathbf{k})-\bar{e}_i(\mathbf{k})\bar{e}_j(\mathbf{k})\Bigg),\quad
e_{ij}^{\times}(\mathbf{k})=\frac{1}{\sqrt{2}}\Bigg(e_i(\mathbf{k})\bar{e}_j(\mathbf{k})+\bar{e}_i(\mathbf{k})e_j(\mathbf{k})\Bigg),
\q
the normalized vectors $e_i(\mathbf{k})$ and $\bar{e}_i(\mathbf{k})$ are orthogonal to each other and to $\mathbf{k}$.
The tensor equation of motion for $h_{ij}$ can be derived straightforwardly from the perturbed Einstein equation up to the second-order. The scalar perturbation couples from tensor perturbation in the second-order equation. The equation for induced gravitational waves with $\Phi_{\mathbf{k}}$ being the source is given by
\e
h_{\mathbf{k}}^{\prime\prime}(\eta)+2\mathcal{H}h_{\mathbf{k}}^\prime(\eta)+c_g^2k^2h_{\mathbf{k}}(\eta)=4S_{\mathbf{k}}(\eta), \label{gw}
\q
where the prime denotes derivative with respect to conformal time, $\mathcal{H}=a^\prime/a=aH$ is the conformal Hubble parameter, and $c_g$ is the speed of scalar induced gravitational waves. The source term is given by
\e
S_{\mathbf{k}}=\int\frac{\mathrm{d}^3q}{(2\pi)^{3/2}}e_{ij}(\mathbf{k})q_iq_j\Bigg(2\Phi_{\mathbf{q}}\Phi_{\mathbf{k-q}}+\frac{4}{3(1+\omega)}\left(\mathcal{H}^{-1}\Phi^{\prime}_{\mathbf{q}}+\Phi_{\mathbf{q}}\right)\left(\mathcal{H}^{-1}\Phi^{\prime}_{\mathbf{k-q}}+\Phi_{\mathbf{k-q}}\right)\Bigg).
\q
We consider the Green's function method as
\e
h_{\mathbf{k}}(\eta)=\frac{4}{a(\eta)}\int^{\eta}{\mathrm{d}}\bar{\eta}G_{\mathbf{k}}(\eta,\bar{\eta})a(\bar{\eta})S_{\mathbf{k}}(\bar{\eta}),
\q
where $G_{\mathbf{k}}(\eta,\bar{\eta})$ satisfies the equation
\e
G_{\mathbf{k}}^{\prime\prime}(\eta,\bar{\eta})+\left(c_g^2k^2-\frac{a^{\prime\prime}(\eta)}{a(\eta)}\right)G_{\mathbf{k}}(\eta,\bar{\eta})=\delta(\eta-\bar{\eta}).
\q
In the Radiation dominated Universe, the solution of the Green's function is
\e
G_{\mathbf{k}}(\eta,\bar{\eta})=\frac{1}{k}\sin[c_gk(\eta-\bar{\eta})].
\q

The power spectrum of scalar induced gravitational waves is defined as
\e
\langle h_{\mathbf{k}}(\eta) h_{\mathbf{k}^{\prime}}(\eta)\rangle=\frac{2\pi^2}{k^3}\delta^{(3)}(\mathbf{k}+\mathbf{k}^{\prime})\mathcal{P}_h(\eta, k),
\q
and the fractional energy density is
\e
\Omega_{\mathrm{GW}}(\eta, k)=\frac{1}{24}\Bigg(\frac{k}{aH}\Bigg)^2\overline{\mathcal{P}_h(\eta, k)}.
\q
After calculation, the power spectrum of scalar induced gravitational waves takes the form
\e
\mathcal{P}_h(\eta, k)=4\int_0^\infty\mathrm{d}v\int_{\vert1-v\vert}^{1+v}\mathrm{d}u\left(\frac{4v^2-(1+v^2-u^2)^2}{4vu}\right)^2I^2(v, u, x)\mathcal{P}_{\zeta}(kv)\mathcal{P}_{\zeta}(ku), \label{ph}
\q
where $\mathcal{P}_{\zeta}(k)$ is the power spectrum of the primordial curvature perturbations, $x\equiv k\eta$, $u=|\mathbf{k}-\tilde{\mathbf{k}}|/k$ and $v=\tilde k/k$. The function $I(v, u, x)$ is defined as
\e
I(v, u, x)=\int_0^x\mathrm{d}\bar{x}\frac{a(\bar{\eta})}{a(\eta)}kG_{\mathbf{k}}(\eta,\bar{\eta})f(v, u, \bar{x}),
\q
where $\bar{x}\equiv k\bar{\eta}$, and $f(v, u, \bar{x})$ comes from the source term
\begin{align}
f(v, u, \bar{x})=&\frac{6(\omega+1)}{3\omega+5}\Phi(v\bar x)\Phi(u\bar x)+\frac{6(1+3\omega)(\omega+1)}{(3\omega+5)^2}\Big(\bar x \partial_{\bar{\eta}}\Phi(v\bar x)\Phi(u\bar x)+\bar x \partial_{\bar{\eta}}\Phi(u\bar x)\Phi(v\bar x)\Big) \nonumber\\
&+\frac{3(1+3\omega)^2(1+\omega)}{(3\omega+5)^2}{\bar x}^2\partial_{\bar{\eta}}\Phi(v\bar x)\partial_{\bar{\eta}}\Phi(u\bar x).
\end{align}
In the radiation-dominated Universe, the results are
\begin{align}
f_{\mathrm{RD}}(v, u, \bar{x})=&\frac{12}{u^3v^3\bar{x}^6}\Big(18uv{\bar{x}}^2\cos\frac{u\bar{x}}{\sqrt{3}}\cos\frac{v\bar{x}}{\sqrt{3}}+(54-6(u^2+v^2){\bar{x}}^2+u^2v^2{\bar{x}}^4)\sin\frac{u\bar{x}}{\sqrt{3}}\sin\frac{v\bar{x}}{\sqrt{3}}\nonumber\\
&+2\sqrt{3}u\bar{x}(v^2{\bar{x}}^2-9)\cos\frac{u\bar{x}}{\sqrt{3}}\sin\frac{v\bar{x}}{\sqrt{3}}+2\sqrt{3}v\bar{x}(u^2{\bar{x}}^2-9)\sin\frac{u\bar{x}}{\sqrt{3}}\cos\frac{v\bar{x}}{\sqrt{3}}\Big),
\end{align}
\begin{align}
I_{\mathrm{RD}}(v, u, x)=&\frac{3}{4u^3v^3x}\Big(-\frac{4}{x^3}\Big(uv(u^2+v^2-3c_g^2)x^3\sin(c_gx)-3c_g(6+(u^2+v^2-3c_g^2)x^2)\sin\frac{ux}{\sqrt{3}}\sin\frac{vx}{\sqrt{3}}\nonumber\\
&-6c_guvx^2\cos\frac{ux}{\sqrt{3}}\cos\frac{vx}{\sqrt{3}}+6\sqrt{3}c_gux\cos\frac{ux}{\sqrt{3}}\sin\frac{vx}{\sqrt{3}}+6\sqrt{3}c_gvx\sin\frac{ux}{\sqrt{3}}\cos\frac{vx}{\sqrt{3}}\Big)\nonumber\\
&+(u^2+v^2-3c_g^2)^2\Big(\sin(c_gx)\Big(\operatorname{Ci}\Big(\Big(c_g-\frac{v-u}{\sqrt{3}}\Big)x\Big)+\operatorname{Ci}\Big(\Big(c_g+\frac{v-u}{\sqrt{3}}\Big)x\Big)\nonumber\\
&-\operatorname{Ci}\Big(\Big(c_g+\frac{v+u}{\sqrt{3}}\Big)x\Big)-\operatorname{Ci}\Big(\left|c_g-\frac{v+u}{\sqrt{3}}\right| x\Big)+\log\left|\frac{3c_g^2-(u+v)^2}{3c_g^2-(u-v)^2}\right|\Big)\nonumber\\
&+\cos(c_gx)\Big(-\operatorname{Si}\Big(\Big(c_g-\frac{v-u}{\sqrt{3}}\Big)x\Big)-\operatorname{Si}\Big(\Big(c_g+\frac{v-u}{\sqrt{3}}\Big)x\Big)\nonumber\\
&+\operatorname{Si}\Big(\Big(c_g+\frac{v+u}{\sqrt{3}}\Big)x\Big)+\operatorname{Si}\Big(\Big(c_g-\frac{v+u}{\sqrt{3}}\Big)x\Big)\Big)\Big).
\end{align}
When we consider the power spectrum of scalar induced gravitational waves today, we can take the limit $x\to\infty$ as
\begin{align}
I_{\mathrm{RD}}(v, u, x\to\infty)=&\frac{3(u^2+v^2-3c_g^2)}{4u^3v^3x}\Big(\sin(c_gx)\Big(-4uv+(u^2+v^2-3c_g^2)\log\left|\frac{3c_g^2-(u+v)^2}{3c_g^2-(u-v)^2}\right|\Big)\nonumber\\
&-\pi\cos(c_gx)(u^2+v^2-3c_g^2)\Theta(u+v-\sqrt{3}c_g)\Big),
\end{align}
where $\Theta(u+v-\sqrt{3}c_g)$ is the Heaviside theta function. The oscillation average is
\begin{align}
\overline{I_{\mathrm{RD}}^2(v, u, x\to\infty)}=&\frac{1}{2}\Bigg(\frac{3(u^2+v^2-3c_g^2)}{4u^3v^3x}\Bigg)^2\Bigg(\Big(-4uv+(u^2+v^2-3c_g^2)\log\left|\frac{3c_g^2-(u+v)^2}{3c_g^2-(u-v)^2}\right|\Big)^2\nonumber\\
&+\pi^2(u^2+v^2-3c_g^2)^2\Theta(u+v-\sqrt{3}c_g)\Bigg).
\end{align}
In the $c_g=1$ case, the scalar induced gravitational waves propagate at the same speed as light \cite{Kohri:2018awv,Lu:2019sti,Li:2022avp,Ananda:2006af,Baumann:2007zm,Inomata:2016rbd,Li:2021uvn}. In the following, we discuss some examples.

\subsection{The monochromatic power spectrum}
\begin{figure}[htb]
\centering
\includegraphics[width=9.9cm]{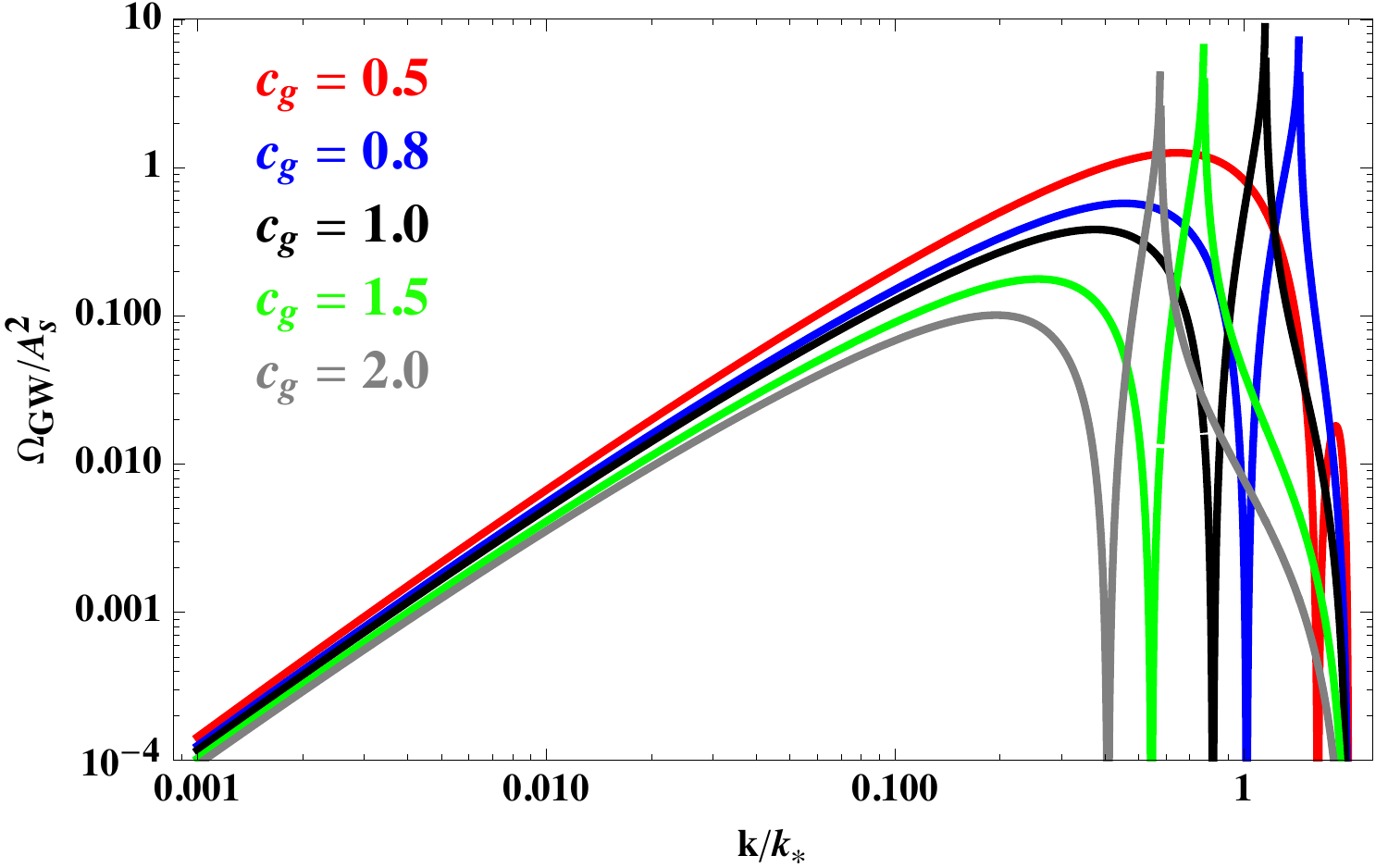}
\caption{The energy density fraction $\Omega_{\mathrm{GW}}$ of the scalar induced gravitational waves in Eq.~(\ref{monochromatic}) from the monochromatic power spectrum. }
\label{figure1}
\end{figure}

The monochromatic curvature perturbations generate a delta-function-type power spectrum \cite{Ananda:2006af,Kohri:2018awv,Lu:2019sti,Inomata:2016rbd}
\e
\mathcal{P}_{\zeta}(k)=A_s\delta\Big(\log{\frac{k}{k_*}}\Big),
\q
where $A_s$ is the scalar amplitude and $k_*$ is the wavenumber at which the delta-function occurs. The fractional energy density becomes
\begin{align}
\Omega_{\mathrm{GW}}(\eta, k)=&\frac{3A_s^2}{64}\Big(\frac{4-{\bar{k}}^2}{4}\Big)^2{\bar{k}}^2\left(3c_g^2{\bar{k}}^2-2\right)^2\Big(\Big(4+(3c_g^2{\bar{k}}^2-2)\log\left|1-\frac{4}{3c_g^2{\bar{k}}^2}\right|\Big)^2\nonumber\\
&+\pi^2(3c_g^2{\bar{k}}^2-2)^2\Theta(2-\sqrt{3}c_g\bar k)\Big)\Theta(2-\bar k),\label{monochromatic}
\end{align}
where $\bar k\equiv{k}/{k_*}$ is the dimensionless wavenumber. In the calculation, only the mode $u=v={\bar k}^{-1}$ contributes to the integration in Eq.~(\ref{ph}). The spectrum vanishes above $k=2k_*$ because no solutions satisfy the energy and momentum conservation. As shown in Fig.~\ref{figure1}, signatures of the speed of scalar induced gravitational waves in the energy density fraction is obvious.

\subsection{The scale invariant power spectrum}
The scale invariant power spectrum is
\e
\mathcal{P}_{\zeta}(k)=A_s,
\q
which is independent of $k$. The fractional energy density becomes
\begin{align}
\Omega_{\mathrm{GW}}(\eta, k)=Q(c_g)A_s^2, \label{scale invariant}
\end{align}
where the overall coefficient $Q(c_g)$ are given in Table.~\ref{table1}.

\begin{table*}[htb]
\newcommand{\tabincell}[2]{\begin{tabular}{@{}#1@{}}#2\end{tabular}}
  \centering
  \resizebox{0.37\textwidth}{!}{
  \begin{tabular}{  c |c| c| c|c|c}
 \hline
 \hline
   $c_g$ & $0.8$ & $1.0$ & $1.5$ & $2.0$ & $2.5$\\
  \hline
  $Q(c_g)$ & $1.1269$ &$0.8211$  &$0.3331$&$0.1564$&$0.0842$\\
  \hline
  \end{tabular}}
  \caption{The overall coefficient $Q(c_g)$ in the energy density fraction of the scalar induced gravitational waves in Eq.~(\ref{scale invariant}) from the scale invariant power spectrum.}
  \label{table1}
\end{table*}

\subsection{The power-law power spectrum}
For a power-law scalar power spectrum
\e
\mathcal{P}_{\zeta}(k)=A_s\(\frac{k}{k_*}\)^{n_s-1},
\q
the power spectrum of scalar induced gravitational waves takes the form
\e
\mathcal{P}_h(\eta, k)=\frac{24Q(c_g, n_s)}{(k\eta)^2}A^2_s\(\frac{k}{k_*}\)^{2(n_s-1)},
\q
where $A_s$ is the scalar amplitude at the pivot scale $k_*=0.05$ Mpc$^{-1}$, $n_s$ is the scalar spectral index and $Q(c_g, n_s)$ is the overall coefficient
\begin{align}
Q(c_g, n_s)=&\frac{1}{12}\int_0^\infty\mathrm{d}v\int_{\vert1-v\vert}^{1+v}\mathrm{d}u \Bigg(\frac{4v^2-(1+v^2-u^2)^2}{4vu}\Bigg)^2\Bigg(\frac{3(u^2+v^2-3c_g^2)}{4u^3v^3}\Bigg)^2\Bigg(uv\Bigg)^{n_s-1} \nonumber\\
&\Bigg(\Big(-4uv+(u^2+v^2-3c_g^2)\log\left|\frac{3c_g^2-(u+v)^2}{3c_g^2-(u-v)^2}\right|\Big)^2+\pi^2(u^2+v^2-3c_g^2)^2\Theta(u+v-\sqrt{3}c_g)\Bigg),
\end{align}
which depends on $c_g$ and $n_s$ as Table.~\ref{table2}. According to Planck18\renewcommand{\thefootnote}{\Roman{footnote}}\footnote{Planck18=TTTEEE+lowE+lensing}+BAO constraint \cite{Planck:2018vyg}, the central value is $n_s=0.9665\pm{0.0038}$.
The fractional energy density becomes
\e
\Omega_{\mathrm{GW}}(\eta, k)=Q(c_g, n_s)A^2_s\(\frac{k}{k_*}\)^{2(n_s-1)},  \label{power law}
\q
which corresponds to the quantity evaluated at late times during the radiation dominated era. If it is evaluated today, the present value of the energy fraction is related to the value in the radiation dominated era
\e
\Omega_{\mathrm{GW}}(\eta_0, k)=\Omega_{r,0}\Omega_{\mathrm{GW}}(\eta_c, k), \label{igwtoday}
\q
where $\Omega_{r,0}=\rho_{r,0}/\rho_0$ is the present value of the energy density fraction of radiation and $\eta_c$ is some time after $\Omega_{\mathrm{GW}}(\eta, k)$ has become constant.

\begin{table}[thb]
\newcommand{\tabincell}[2]{\begin{tabular}{@{}#1@{}}#2\end{tabular}}
  \centering
  \resizebox{0.65\textwidth}{!}{
  \begin{tabular}{  c |c| c| c|c|c|c|c|c|c}
 \hline
 \hline
   \diagbox{$c_g$}{$n_s$}&$0.4$ &$0.6$ & $0.8$&$1.0$& $1.2$& $1.4$&$1.6$& $1.8$& $0.9665$\\
  \hline
  $0.8$ & $1.356$ &$1.227$  &$1.149$&$1.127$&$1.191$&$1.395$&$1.888$&$3.126$&$1.127$\\
  \hline
   $1.0$ & $0.8182$ &$0.7971$  &$0.7959$&$0.8211$&$0.8988$&$1.074$&$1.470$&$2.478$&$0.8151$\\
  \hline
   $1.5$ & $0.2165$ &$0.2445$  &$0.2817$&$0.3331$&$0.4129$&$0.5509$&$0.8320$&$1.539$&$0.3233$\\
  \hline
   $2.0$ & $0.0741$ &$0.0930$  &$0.1190$&$0.1564$&$0.2148$&$0.3164$&$0.5252$&$1.065$&$0.1491$\\
  \hline
   $2.5$ & $0.0312$ &$0.0424$  &$0.0590$&$0.0842$&$0.1257$&$0.2009$&$0.3612$&$0.7925$&$0.0792$\\
  \hline
  \end{tabular}}
  \caption{The overall coefficient $Q(c_g, n_s)$ in the energy density fraction of the scalar induced gravitational waves in Eq.~(\ref{power law}) from the power-law power spectrum.}
  \label{table2}
\end{table}

Then, we characterize the scalar fluctuation spectrum in terms of the spectral index $n_s$ and its first derivatives with respect to $\ln k$ as \cite{Li:2022avp}
\e
\mathcal{P}_{\zeta}(k)=A_s\(\frac{k}{k_*}\)^{n_s-1+\frac{1}{2}\alpha_s\ln(k/k_*)},
\q
where $\alpha_s\equiv\mathrm{d} n_s/\mathrm{d}\ln k$ is the running of the spectral index. The power spectrum of scalar induced gravitational waves takes the form
\e
\mathcal{P}_h(\eta, k)=\frac{24Q(c_g, n_s, \alpha_s, k)}{(k\eta)^2}A^2_s\(\frac{k}{k_*}\)^{2\[{n_s-1+\frac{1}{2}\alpha_s\ln(k/k_*)}\]},
\q
where the overall coefficient is
\begin{align}
Q(c_g, n_s, \alpha_s, k)=&\frac{1}{12}\int_0^\infty\mathrm{d}v\int_{\vert1-v\vert}^{1+v}\mathrm{d}u \Bigg(\frac{4v^2-(1+v^2-u^2)^2}{4vu}\Bigg)^2\Bigg(\frac{3(u^2+v^2-3c_g^2)}{4u^3v^3}\Bigg)^2 \nonumber\\
&\Bigg(\Big(-4uv+(u^2+v^2-3c_g^2)\log\left|\frac{3c_g^2-(u+v)^2}{3c_g^2-(u-v)^2}\right|\Big)^2+\pi^2(u^2+v^2-3c_g^2)^2\Theta(u+v-\sqrt{3}c_g)\Bigg) \nonumber\\
&\Bigg(\frac{k}{k_*}\Bigg)^{\frac{1}{2}\alpha_s\ln(uv)}\Bigg(uv\Bigg)^{n_s-1+\frac{1}{2}\alpha_s\ln(k/k_*)}v^{\frac{1}{2}\alpha_s\ln{v}}u^{\frac{1}{2}\alpha_s\ln{u}},
\end{align}
which depends on $c_g$, $n_s$, $\alpha_s$ and $k$. According to Planck18+BAO constraints, the central values are $n_s=0.9659\pm0.0040$ and $\alpha_s=-0.0041\pm0.0067$. We fix $n_s$ and $\alpha_s$, and obtain some overall coefficients $Q(c_g, n_s, \alpha_s, k)$ in Table.~\ref{table3}.
The fractional energy density becomes
\e
\Omega_{\mathrm{GW}}(\eta, k)=Q(c_g, n_s, \alpha_s, k)A^2_s\(\frac{k}{k_*}\)^{2\[{n_s-1+\frac{1}{2}\alpha_s\ln(k/k_*)}\]}.  \label{power law1}
\q

\begin{table*}[htb]
\newcommand{\tabincell}[2]{\begin{tabular}{@{}#1@{}}#2\end{tabular}}
  \centering
  \resizebox{0.58\textwidth}{!}{
  \begin{tabular}{  c |c| c| c|c|c}
 \hline
 \hline
   \diagbox{$c_g$}{$k/k_*$} & $4.09*10^6$&$8.18*10^6$&$2.04*10^7$&$1.29*10^{13}$&$5.16*10^{17}$\\
  \hline
  $0.8$ & $1.127$ &$1.128$  &$1.128$&$1.137$&$1.147$\\
  \hline
   $1.0$ & $0.8038$ &$0.8035$  &$0.8030$&$0.7975$&$0.7949$\\
  \hline
   $1.5$ & $0.3055$ &$0.3048$  &$0.3039$&$0.2907$&$0.2811$\\
  \hline
   $2.0$ & $0.1362$ &$0.1356$  &$0.1350$&$0.1254$&$0.1185$\\
  \hline
   $2.5$ & $0.0704$ &$0.0700$  &$0.0695$&$0.0632$&$0.0586$\\
  \hline
  \end{tabular}}
  \caption{The overall coefficient $Q(c_g, n_s, \alpha_s, k)$ in the energy density fraction of the scalar induced gravitational waves in Eq.~(\ref{power law1}) from the power-law power spectrum.}
  \label{table3}
\end{table*}

Furthermore, we can characterize the scalar fluctuation spectrum in terms of the spectral index $n_s$ and its first two derivatives with respect to $\ln k$
\e
\mathcal{P}_{\zeta}(k)=A_s\(\frac{k}{k_*}\)^{n_s-1+\frac{1}{2}\alpha_s\ln(k/k_*)+\frac{1}{6}\beta_s(\ln(k/k_*))^2},
\q
where $\beta_s\equiv{\mathrm{d}^2n_s}/{\mathrm{d}\ln k^2}$ is the running of the running of the spectral index. The power spectrum of scalar induced gravitational waves takes the form
\e
\mathcal{P}_h(\eta, k)=\frac{24Q(c_g, n_s, \alpha_s, \beta_s, k)}{(k\eta)^2}A^2_s\(\frac{k}{k_*}\)^{2\[{n_s-1+\frac{1}{2}\alpha_s\ln(k/k_*)+\frac{1}{6}\beta_s(\ln(k/k_*))^2}\]},
\q
where the overall coefficient is
\begin{align}
Q(c_g, n_s, \alpha_s, \beta_s, k)=&\frac{1}{12}\int_0^\infty\mathrm{d}v\int_{\vert1-v\vert}^{1+v}\mathrm{d}u \Bigg(\frac{4v^2-(1+v^2-u^2)^2}{4vu}\Bigg)^2 \Bigg(\frac{3(u^2+v^2-3c_g^2)}{4u^3v^3}\Bigg)^2 \nonumber\\
&\Bigg(\Big(-4uv+(u^2+v^2-3c_g^2)\log\left|\frac{3c_g^2-(u+v)^2}{3c_g^2-(u-v)^2}\right|\Big)^2+\pi^2(u^2+v^2-3c_g^2)^2\Theta(u+v-\sqrt{3}c_g)\Bigg) \nonumber\\
&\Bigg(\frac{k}{k_*}\Bigg)^{\frac{1}{2}\alpha_s\ln(uv)+\frac{1}{6}\beta_s\Big((\ln{v})^2+2\ln{v}\ln(k/k_*)+(\ln{u})^2+2\ln{u}\ln(k/k_*)\Big)}\Bigg(uv\Bigg)^{n_s-1+\frac{1}{2}\alpha_s\ln(k/k_*)+\frac{1}{6}\beta_s(\ln(k/k_*))^2}\nonumber\\
&v^{\frac{1}{2}\alpha_s\ln{v}+\frac{1}{6}\beta_s\Big((\ln{v})^2+2\ln{v}\ln(k/k_*)\Big)}u^{\frac{1}{2}\alpha_s\ln{u}+\frac{1}{6}\beta_s\Big((\ln{u})^2+2\ln{u}\ln(k/k_*)\Big)},
\end{align}
which depends on $c_g$, $n_s$, $\alpha_s$, $\beta_s$ and $k$. According to Planck18+BAO constraints, the central values are $n_s=0.9647\pm0.0043$, $\alpha_s=0.009\pm0.012$ and $\beta_s=0.0011\pm0.0099$. We fix $n_s$, $\alpha_s$ and $\beta_s$, and obtain some overall coefficients $Q(c_g, n_s, \alpha_s, \beta_s, k)$ in Table.~\ref{table4}.
The fractional energy density becomes
\e
\Omega_{\mathrm{GW}}(\eta, k)=Q(c_g, n_s, \alpha_s, \beta_s, k)A^2_s\(\frac{k}{k_*}\)^{2\[{n_s-1+\frac{1}{2}\alpha_s\ln(k/k_*)+\frac{1}{6}\beta_s(\ln(k/k_*))^2}\]}.  \label{power law2}
\q

\begin{table*}[htb]
\newcommand{\tabincell}[2]{\begin{tabular}{@{}#1@{}}#2\end{tabular}}
  \centering
  \resizebox{0.38\textwidth}{!}{
  \begin{tabular}{  c |c| c| c}
 \hline
 \hline
   \diagbox{$c_g$}{$k/k_*$} &  $4.09*10^6$ & $8.18*10^6$ & $2.04*10^7$\\
  \hline
  $0.9$ & $2618.46$ &$968.76$  &$841.32$\\
  \hline
   $1.0$ & $224.44$ &$193.02$  &$156.83$\\
  \hline
   $1.2$ & $826.68$ &$680.79$  &$520.46$\\
  \hline
   $1.6$ & $1356.45$ &$1101.47$  &$7950.39$\\
  \hline
   $2.0$ & $2.35*10^{6}$ &$2.02*10^{6}$  &$1.63*10^{6}$\\
  \hline
  \end{tabular}}
  \caption{The overall coefficient $Q(c_g, n_s, \alpha_s, \beta_s, k)$ in the energy density fraction of the scalar induced gravitational waves in Eq.~(\ref{power law2}) from the power-law power spectrum.}
  \label{table4}
\end{table*}

\section{measuring the speed of scalar induced gravitational waves from observations}
\begin{figure}[htb]
\centering
\includegraphics[width=18cm]{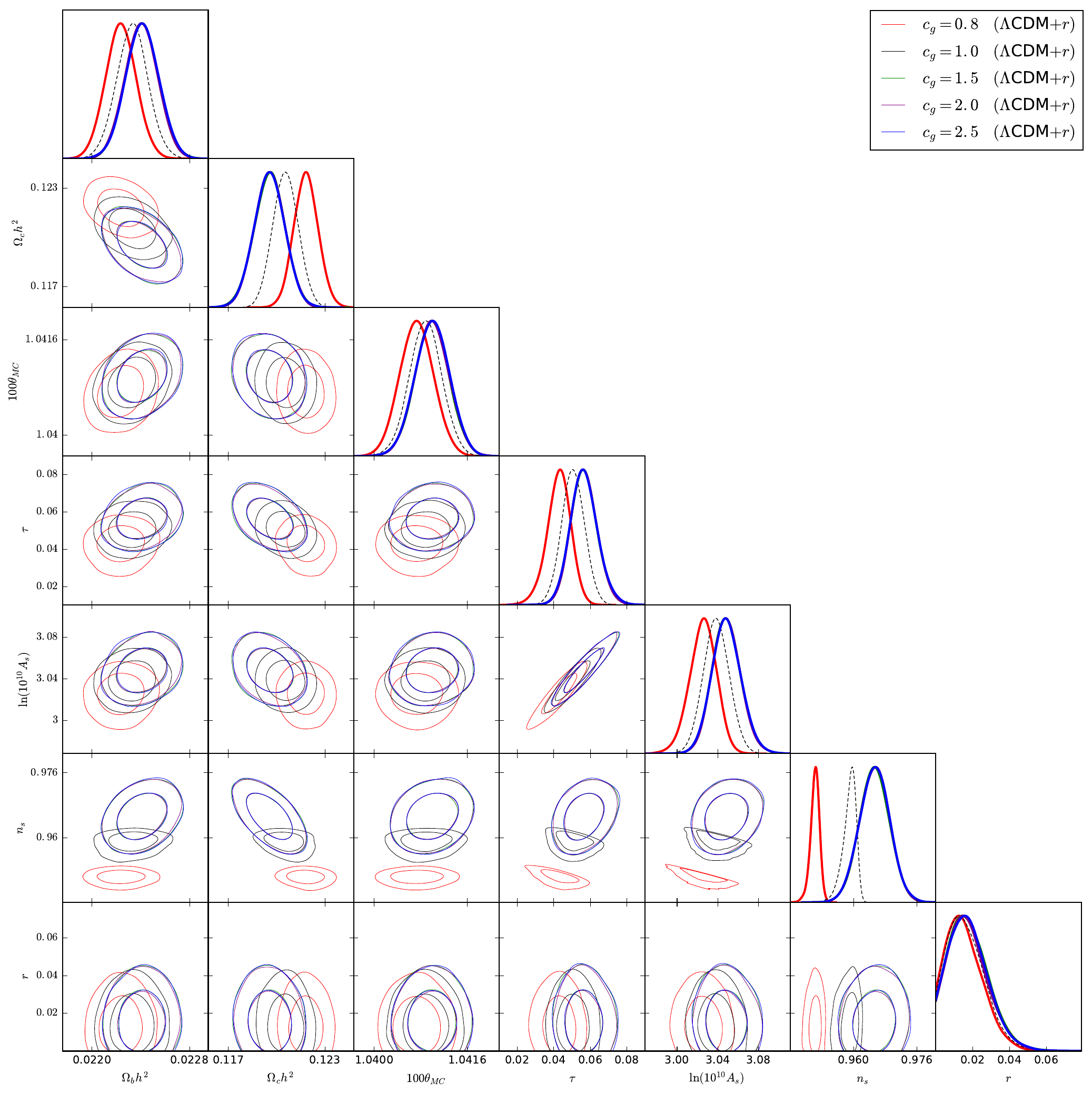}
\caption{The contour plots and the likelihood distributions for the cosmological parameters in the $\Lambda$CDM+$r$ model at the $68\%$ and $95\%$ CL from the combination of CMB+BAO+SKA in the cases $c_g=0.8$, $c_g=1.0$, $c_g=1.5$, $c_g=2.0$ and $c_g=2.5$, respectively. The dashed black lines in the likelihood distributions are belong to $c_g=1.0$ case.}
\label{figure2}
\end{figure}

We use the publicly available codes Cosmomc \cite{Lewis:2002ah} to constrain the scalar induced gravitational waves and evaluate the signatures of the speed. In the standard $\Lambda$CDM model, the six parameters are the baryon density parameter $\Omega_b h^2$, the cold dark matter density $\Omega_c h^2$, the angular size of the horizon at the last scattering surface $\theta_\text{MC}$, the optical depth $\tau$, the scalar amplitude $A_s$ and the scalar spectral index $n_s$. Usually we introduce a new parameter, namely the tensor-to-scalar ratio $r$, to quantify the tensor amplitude $A_t$ compared to the scalar amplitude $A_s$
at the pivot scale:
\e
r\equiv\frac{A_t}{A_s}.
\q
Here, we consider the power-law spectrum first. We extend the standard $\Lambda$CDM model by adding the tensor-to-scalar ratio $r$, and constrain these seven parameters from the combination of CMB\renewcommand{\thefootnote}{\Roman{footnote}}\footnote{CMB=Planck18+BK18}+BAO+SKA in the cases $c_g=0.8$, $c_g=1.0$, $c_g=1.5$, $c_g=2.0$ and $c_g=2.5$, respectively. Our numerical results are given in Table.~\ref{table5} and Fig.~\ref{figure2}.

\begin{table*}[htb]
\newcommand{\tabincell}[2]{\begin{tabular}{@{}#1@{}}#2\end{tabular}}
  \centering
   \resizebox{1.0\textwidth}{!}{
  \begin{tabular}{  c |c| c| c|c|c}
  \hline
  \hline
  Parameter & \tabincell{c} {$c_g=0.8$} & \tabincell{c}{$c_g=1.0$} & \tabincell{c}{$c_g=1.5$}& \tabincell{c}{$c_g=2.0$}& \tabincell{c}{$c_g=2.5$}\\
  \hline
  $\Omega_bh^2$ & $0.02224^{+0.00012}_{-0.00013}$&$0.02233\pm0.00013$&$0.02241\pm0.00013$&$0.02241\pm0.00013$&$0.02241\pm0.00014$\\
  $\Omega_ch^2$ &$0.12183^{+0.00075}_{-0.00076}$&$0.12056\pm{0.00079}$&$0.11956^{+0.00094}_{-0.00095}$&$0.11956\pm0.00093$&$0.11955^{+0.00094}_{-0.00095}$\\
  $100\theta_{\mathrm{MC}}$ & $1.04073\pm0.00029$&$1.04088^{+0.00027}_{-0.00028}$&$1.04099^{+0.00028}_{-0.00029}$&$1.04099^{+0.00030}_{-0.00029}$&$1.04099^{+0.00030}_{-0.00029}$\\
  $\tau$ &  $0.0428^{+0.0070}_{-0.0061}$&$0.0505^{+0.0063}_{-0.0062}$&$0.0565^{+0.0070}_{-0.0078}$&$0.0565^{+0.0068}_{-0.0077}$&$0.0566^{+0.0070}_{-0.0079}$\\
  $\ln\(10^{10}A_s\)$  & $3.025^{+0.014}_{-0.013}$&$3.038\pm0.013$&$3.048^{+0.014}_{-0.015}$&$3.048\pm0.014$&$3.049^{+0.014}_{-0.015}$\\
  $n_s$ & $0.9503^{+0.0013}_{-0.0010}$&$0.9589^{+0.0021}_{-0.0011}$&$0.9652\pm0.0037$&$0.9653^{+0.0037}_{-0.0038}$&$0.9653^{+0.0037}_{-0.0038}$\\
  $r_{0.05}$  ($95\%$ CL) &$<0.034$ &$<0.035$&$<0.037$&$<0.037$&$<0.037$\\
  \hline
  \end{tabular}}
  \caption{The $68\%$ limits on the cosmological parameters in the $\Lambda$CDM+$r$ model from the combination of CMB+BAO+SKA in the cases $c_g=0.8$, $c_g=1.0$, $c_g=1.5$, $c_g=2.0$ and $c_g=2.5$, respectively.}
  \label{table5}
\end{table*}

In the $\Lambda$CDM+$r$ model, we see that the constraints on the cosmological parameters are affected by the speed of scalar induced gravitational waves. The subluminal case and superluminal cases are much different from the light one. After comparing $c_g=0.8$, $c_g=1.0$ and $c_g=1.5$ cases, we find that the mean values of the scalar amplitude $A_s$ and spectral index $n_s$ shift to lower values when the speed of scalar induced gravitational waves become smaller. Although the differences between the subluminal case, the light case and the superluminal cases are obvious, but it is hard to distinguish three superluminal cases $c_g=1.5$, $c_g=2.0$ and $c_g=2.5$ in the $\Lambda$CDM+$r$ model.

\begin{figure}[htb]
\centering
\includegraphics[width=18cm]{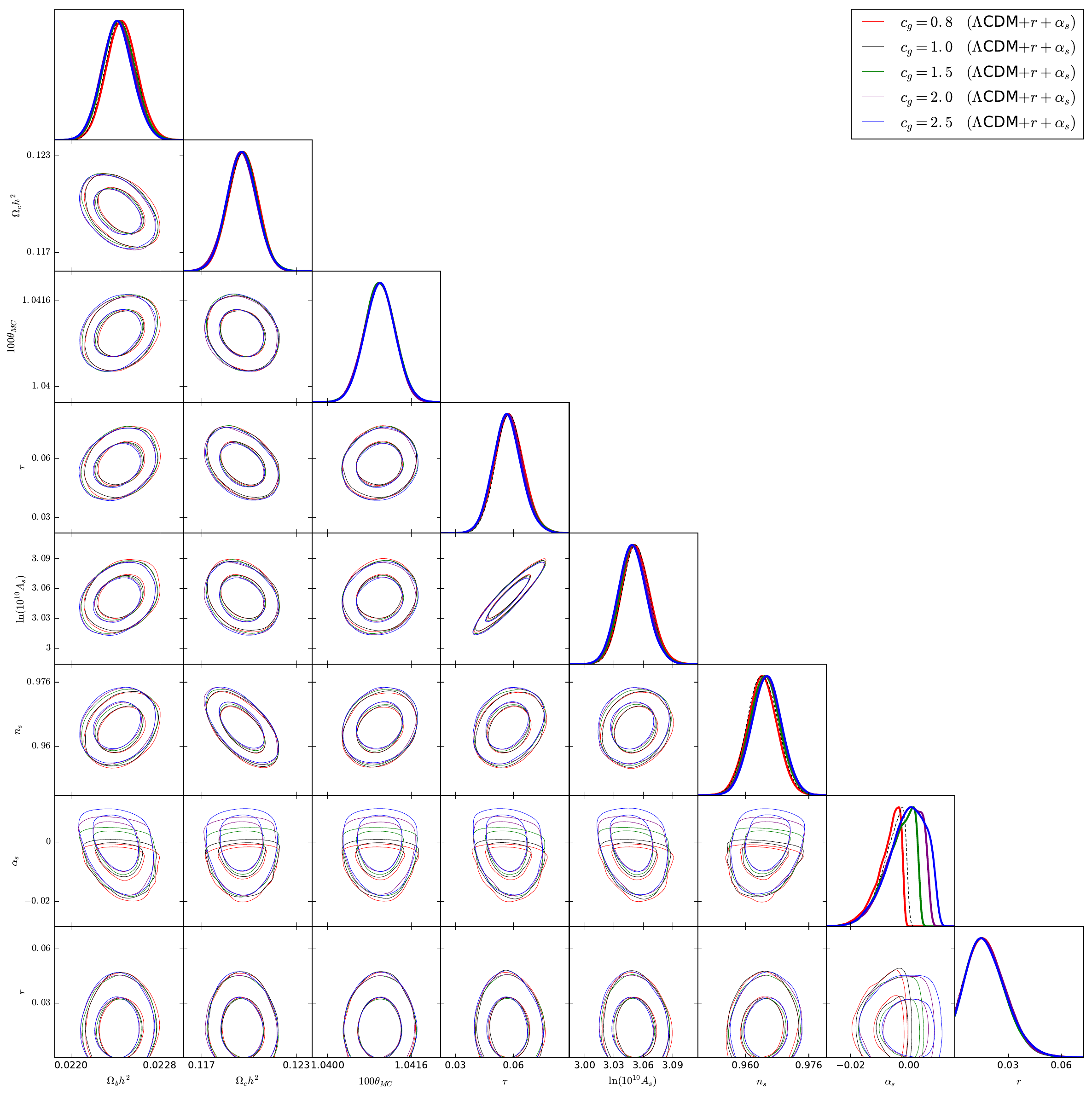}
\caption{The contour plots and the likelihood distributions for the cosmological parameters in the $\Lambda$CDM+$r$+$\alpha_s$ model at the $68\%$ and $95\%$ CL from the combination of CMB+BAO+SKA in the cases $c_g=0.8$, $c_g=1.0$, $c_g=1.5$, $c_g=2.0$ and $c_g=2.5$, respectively. The dashed black lines in the likelihood distributions are belong to $c_g=1.0$ case.}
\label{figure3}
\end{figure}

\begin{figure}[htb]
\centering
\includegraphics[width=18cm]{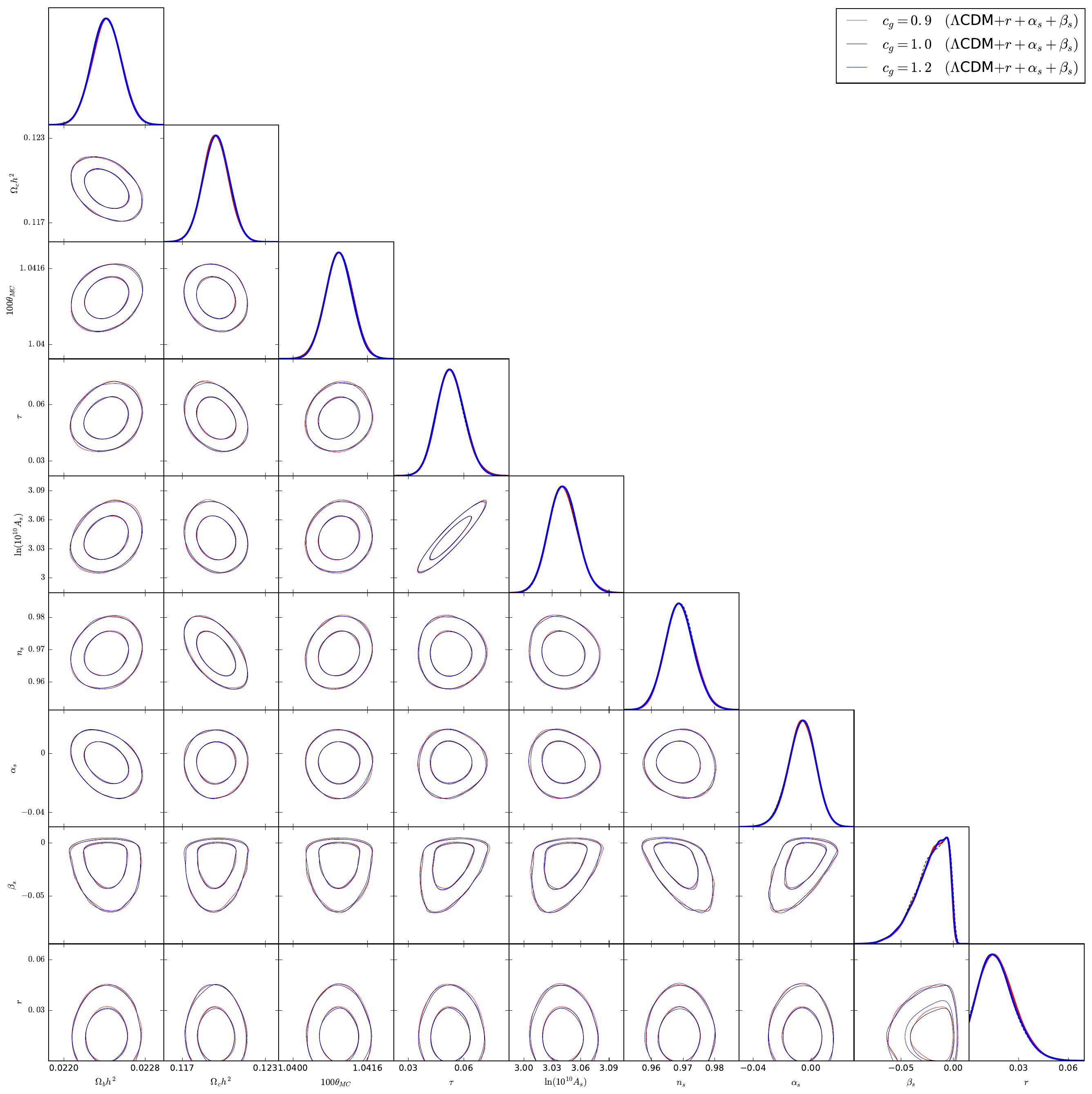}
\caption{The contour plots and the likelihood distributions for the cosmological parameters in the $\Lambda$CDM+$r$+$\alpha_s$+$\beta_s$ model at the $68\%$ and $95\%$ CL from the combination of CMB+BAO+FAST in the cases $c_g=0.9$, $c_g=1.0$ and $c_g=1.2$, respectively. The dashed black lines in the likelihood distributions are belong to $c_g=1.0$ case.}
\label{figure4}
\end{figure}

Then, we add the running of the spectral index $\alpha_s$ and the running of the running of the spectral index $\beta_s$ into the $\Lambda$CDM+$r$ model. We investigate the $\Lambda$CDM+$r$+$\alpha_s$ model from the combination of CMB+BAO+SKA in the cases $c_g=0.8$, $c_g=1.0$, $c_g=1.5$, $c_g=2.0$ and $c_g=2.5$, respectively. Also, we explore the $\Lambda$CDM+$r$+$\alpha_s$+$\beta_s$ model from the combination of CMB+BAO+FAST in the cases $c_g=0.9$, $c_g=1.0$ and $c_g=1.2$, respectively. Our numerical results are given in Table.~\ref{table6}, Table.~\ref{table7} and Fig.~\ref{figure3} to Fig.~\ref{figure4}.

\begin{table*}[htb]
\newcommand{\tabincell}[2]{\begin{tabular}{@{}#1@{}}#2\end{tabular}}
  \centering
   \resizebox{1.0\textwidth}{!}{
  \begin{tabular}{  c |c| c| c|c|c}
  \hline
  \hline
  Parameter & \tabincell{c} {$c_g=0.8$} & \tabincell{c}{$c_g=1.0$} & \tabincell{c}{$c_g=1.5$}& \tabincell{c}{$c_g=2.0$}& \tabincell{c}{$c_g=2.5$}\\
  \hline
  $\Omega_bh^2$ & $0.02246\pm0.00014$   &$0.02245\pm0.00014$ &$0.02243\pm0.00014$ &$0.02242\pm0.00014$ & $0.02242\pm0.00014$\\
  $\Omega_ch^2$ &$0.11960\pm0.00093$    &$0.11960_{-0.00094}^{+0.00092}$ & $0.11956\pm0.00093$ &$0.11950^{+0.00094}_{-0.00093}$&$0.11953\pm0.00093$\\
  $100\theta_{\mathrm{MC}}$ & $1.04099\pm0.00029$      &$1.04100_{-0.00028}^{+0.00029}$& $1.04100^{+0.00030}_{-0.00029}$ & $1.04099\pm0.00029$&$1.04100\pm0.00029$\\
  $\tau$ &  $0.0577\pm0.0074$  &$0.0578^{+0.0070}_{-0.0076}$ &$0.0572^{+0.0071}_{-0.0080}$ & $0.0571\pm0.0073$&$0.0567^{+0.0072}_{-0.0073}$\\
  $\ln\(10^{10}A_s\)$  & $3.052_{-0.016}^{+0.014}$     &$3.052\pm0.014$ & $3.051^{+0.014}_{-0.016}$ & $3.050\pm0.015$&$3.049^{+0.014}_{-0.015}$\\
  $n_s$ & $0.9641_{-0.0037}^{+0.0038}$   &$0.9643\pm{0.0038}$ & $0.9648\pm0.0038$&$0.9652^{+0.0038}_{-0.0039}$&$0.9653\pm0.0038$\\
  $\alpha_s$ &$-0.0074_{-0.0022}^{+0.0055}$ &$-0.0062_{-0.0023}^{+0.0056}$ & $-0.0037^{+0.0070}_{-0.0031}$&$-0.0018^{+0.0078}_{-0.0038}$& $-0.0008^{+0.0083}_{-0.0047}$\\
  $r_{0.05}$  ($95\%$ CL) &$<0.038$   &$<0.039$ & $<0.037$ & $<0.038$&$<0.038$\\
  \hline
  \end{tabular}}
  \caption{The $68\%$ limits on the cosmological parameters in the $\Lambda$CDM+$r$+$\alpha_s$ model from the combination of CMB+BAO+SKA in the cases $c_g=0.8$, $c_g=1.0$, $c_g=1.5$, $c_g=2.0$ and $c_g=2.5$, respectively.}
  \label{table6}
\end{table*}

\begin{table*}[htb]
\newcommand{\tabincell}[2]{\begin{tabular}{@{}#1@{}}#2\end{tabular}}
  \centering
   \resizebox{0.66\textwidth}{!}{
  \begin{tabular}{  c |c| c| c}
  \hline
  \hline
  Parameter & \tabincell{c} {$c_g=0.9$} & \tabincell{c}{$c_g=1.0$} & \tabincell{c}{$c_g=1.2$}\\
  \hline
  $\Omega_bh^2$ & $0.02242\pm0.00014$   &$0.02242\pm0.00014$ &$0.02242\pm0.00014$ \\
  $\Omega_ch^2$ &$0.11941_{-0.00091}^{+0.00092}$ &$0.11941\pm0.00092$ & $0.11944^{+0.00094}_{-0.00093}$ \\
  $100\theta_{\mathrm{MC}}$ & $1.04099\pm0.00029$ &$1.04099\pm0.00029$ & $1.04099^{+0.00028}_{-0.00029}$\\
  $\tau$ &  $0.0529_{-0.0080}^{+0.0073}$  &$0.0530^{+0.0069}_{-0.0081}$ &$0.0529^{+0.0073}_{-0.0080}$ \\
  $\ln\(10^{10}A_s\)$  & $3.041_{-0.016}^{+0.015}$     &$3.042^{+0.014}_{-0.016}$ & $3.041\pm0.015$\\
  $n_s$ & $0.9690_{-0.0049}^{+0.0045}$   &$0.9689^{+0.0045}_{-0.0046}$ & $0.9689\pm0.0045$\\
  $\alpha_s$ &$-0.0063_{-0.009}^{+0.010}$ &$-0.0064_{-0.009}^{+0.010}$ & $-0.0063^{+0.010}_{-0.009}$\\
  $\beta_s$ & $-0.0219_{-0.009}^{+0.020}$&$-0.0217^{+0.021}_{-0.010}$ & $-0.0217^{+0.020}_{-0.009}$\\
  $r_{0.05}$  ($95\%$ CL) &$<0.037$   &$<0.037$ & $<0.037$ \\
  \hline
  \end{tabular}}
  \caption{The $68\%$ limits on the cosmological parameters in the $\Lambda$CDM+$r$+$\alpha_s$+$\beta_s$ model from the combination of CMB+BAO+FAST in the cases $c_g=0.9$, $c_g=1.0$ and $c_g=1.2$, respectively.}
  \label{table7}
\end{table*}

In the $\Lambda$CDM+$r$+$\alpha_s$ model, the signatures of speed of scalar induced gravitational waves are still obvious. When we consider the running of the spectral index $\alpha_s$, the index factor becomes more sensitive and the scalar amplitude becomes less sensitive to the speed of scalar induced gravitational waves. In Fig.~\ref{figure3}, the mean values of the running of the spectral index $\alpha_s$ shift to lower values when the speed of scalar induced gravitational waves become smaller. The differences between the subluminal case, the light case and the superluminal cases are apparent. Meanwhile, it is easy to distinguish three superluminal cases $c_g=1.5$, $c_g=2.0$ and $c_g=2.5$ in the $\Lambda$CDM+$r$+$\alpha_s$ model. In the $\Lambda$CDM+$r$+$\alpha_s$+$\beta_s$ model, the cosmological parameters are less sensitive to the speed of scalar induced gravitational waves. The mean values almost maintain the same values except the running of the running of the spectral index $\beta_s$. It is hard to distinguish the subluminal case, the light case and the superluminal case in the $\Lambda$CDM+$r$+$\alpha_s$+$\beta_s$ model. From Fig.~\ref{figure2} to Fig.~\ref{figure4}, the speed of scalar induced gravitational waves changes the contours and likelihoods from the red ones to the black ones, blue ones which corresponding to the subluminal case, the light case and the superluminal case.

\section{summary}
In this paper, we investigate the scalar induced gravitational waves which propagate with a speed different from the speed of light. First, we analytically calculate the expression of the power spectrum of the scalar induced gravitational waves which is based on the speed and the spectrum of the primordial curvature perturbations. Then, we discuss several scalar power spectra and obtain corresponding fractional energy density, such as the monochromatic power spectrum, the scale invariant power spectrum and the power-law power spectrum. Finally, we constrain the scalar induced gravitational waves and evaluate the signatures of the speed from the combination of CMB+BAO and gravitational waves observations. The numerical results are obvious to reveal the influence of speed of scalar induced gravitational waves.

\noindent {\bf Acknowledgments}.
This work is supported by Natural Science Foundation of Shandong Province (grant No. ZR2021QA073) and Research Start-up Fund of QUST (grant No. 1203043003587).



\end{document}